\documentclass{Interspeech}

\interspeechcameraready 

\title{Open-Set Source Tracing of Audio Deepfake Systems}

\author[affiliation={}]{Nicholas}{Klein}
\author[affiliation={}]{Hemlata}{Tak}
\author[affiliation={}]{Elie}{Khoury}

\affiliation[nocounter]{Speech Research}{Pindrop}{USA}
\email{nklein@pindrop.com, Hemlata.Tak@pindrop.com, ekhoury@pindrop.com}

\usepackage{comment}
\usepackage{tablefootnote}

\usepackage{footnote}

\usepackage{url,cite}
\usepackage{hyperref}
\usepackage[dvipsnames]{xcolor}
\hypersetup{
    colorlinks=true,
    linkcolor=Blue,
    urlcolor=Blue,
    citecolor=BrickRed
}
\newcommand{\newpara}[1]{\vspace{8pt}\noindent\textbf{#1}}
\usepackage{float}
\usepackage{placeins}

\usepackage{booktabs}
\usepackage{cite}

\copyrightnotice{\footnotesize \copyright 2025 Pindrop Security, Inc. This work is licensed under \href{https://creativecommons.org/licenses/by-nc-nd/4.0/legalcode.en}{CC BY-NC-ND 4.0}.}

\begin{document}

\maketitle

\begin{abstract}
Existing research on source tracing of audio deepfake systems has focused primarily on the closed-set scenario, while studies that evaluate open-set performance are limited to a small number of unseen systems. Due to the large number of emerging audio deepfake systems, robust open-set source tracing is critical. We leverage the protocol of the Interspeech 2025 special session on source tracing to evaluate methods for improving open-set source tracing performance. We introduce a novel adaptation to the energy score for out-of-distribution (OOD) detection, softmax energy (SME). We find that replacing the typical temperature-scaled energy score with SME provides a relative average improvement of 31\% in the standard FPR95 (false positive rate at true positive rate of 95\%) measure. We further explore SME-guided training as well as copy synthesis, codec, and reverberation augmentations, yielding an FPR95 of 8.3\%.
\end{abstract}

\section{Introduction}
\thispagestyle{preprint}
In the field of audio deepfake detection, efforts to decipher the provenance of manipulated audio have been gaining attention. This task, known as source tracing or source attribution, holds importance in many sectors such as intellectual property (IP) protection, digital forensics, and enhancing trust in audio deepfake detection systems. Existing work has devised the source tracing task in different ways. The \textbf{choice of granularity} has varied from the model level~\cite{xie2024generalized,muller2022attacker, xie2025neural,borrelli2021synthetic, phukan2024investigating,yi2023add,lu2023detecting, zeng2023deepfake, wang2023npu, tian2023deepfake, qin2023speaker} to higher level components of the model's architecture such as the acoustic model and vocoder~\cite{yan2022initial, zhu2022source, kleinsource, zhang2024distinguishing}. \cite{mishraa2025towards} combines both approaches, leveraging predictions of high level model components to predict the specific source model. The \textbf{use of the genuine class} varied between studies: ~\cite{xie2024generalized, xie2025neural, kleinsource, zhu2022source, yi2023add, lu2023detecting, zeng2023deepfake, wang2023npu, tian2023deepfake, qin2023speaker, muller2022attacker, borrelli2021synthetic, phukan2024investigating} include it among the source models, while~\cite{yan2022initial, mishraa2025towards, zhang2024distinguishing} conceptualize source tracing as a downstream task of deepfake detection and thus focus on separating synthetic speech only. 
A key challenge in building secure trustworthy AI systems is ensuring that deep neural network models can reliably recognize what they do not know, specifically their ability to detect \textbf{out-of-distribution} (OOD) data, where model confidence should naturally be low. Existing source tracing work has considered a limited number of OOD classes. In the ADD2023 challenge~\cite{yi2023add}, Track~3, which focused on source tracing, included a single held out OOD class. Similarly, \cite{xie2024generalized} studied open-set source tracing with one held out OOD class.~\cite{muller2022attacker} found encouraging results clustering samples from four unseen systems. More recently, \cite{xie2025neural} included five OOD classes in their studies.
However, given the ever increasing number of emerging audio deepfake systems, it is imperative that source tracing systems remain robust in an OOD scenario where the number of OOD classes is much larger. Hence, to foster the development of robust source tracing systems in the open-set scenario, Interspeech 2025 organized a special session on \textbf{Source tracing: The origins of synthetic or manipulated speech}. 

In the context of this special session, our paper explores methods to improve the performance of open-set source tracing. 
We use the MLAAD source tracing protocol~\cite{UsingMLAADforSourceTracing} provided by this session to facilitate reproducible results. In particular, this protocol emphasizes the open-set condition, as the evaluation set includes 43 unseen spoof systems. Our main contributions are as follows.
We introduce a novel adaptation to the energy score for OOD detection, softmax energy (SME). We find that replacing the typical temperature-scaled energy score with SME provides a relative improvement in average FPR95 of 31\% across our experiments. We explore SME-guided training, leveraging OODs from ASVspoof~5~\cite{wang2025asvspoof} and copy synthesis, which when paired with codec and reverberation augmentations leads to 8.3\% FPR95. Our best model that does not use auxiliary data outperforms the reference system provided by the special session by an absolute margin of 52\% unweighted EER.

\section{Related Work on OOD Detection}
Many approaches have been proposed to produce scores indicating the probability that a given sample is in-distribution (ID) \emph{versus} OOD. \cite{zhang2023improving}~introduced Energy-based Open-World Softmax which adds an output logit to estimate the model's uncertainty. However, this approach requires modifying the model architecture and training procedure. Thus, we favor exploring methods that can be used with any existing model. \cite{lu2023detecting}~explored the k-Nearest Neighbors (kNN) distance metric which assumes that the k-nearest neighbor of a given sample will be closer for ID data than for OOD data for an appropriately selected k.
~\cite{lee2018simple} also explored a scoring method based on features, leveraging Mahalanobis distance. However, a recent study \cite{xie2025neural} found that Mahalanobis and kNN distance metrics underperform compared to the logits-based methods, Maximum Softmax Probability (MSP)~\cite{hendrycks2022baseline} and energy~\cite{liu2020energy}. MSP is simple but effective, assuming that ID samples will produce larger maximum softmax probabilities than OOD samples. However, MSP has been found to suffer from overconfidence on OOD data. \cite{liang2018enhancing}~aimed to address this with ODIN which applies temperature scaling and input perturbations, and~\cite{liu2020energy} proposed an energy score
which outperforms MSP on a number of benchmarks without requiring tuning of the temperature parameter. Given its efficacy and its ability to be applied to many existing models, the energy score is a popular approach for OOD detection.

However, our work finds that the energy score performance is compromised when applying it to bounded logits produced by models using Large Margin Cosine Loss (LMCL)~\cite{wang2018cosface} which has shown promise in the closed-set source tracing task~\cite{kleinsource}. Thus, we propose a novel adaptation to the energy score, softmax energy (SME), and explore its utility in providing OOD detection scores and guiding training with auxiliary OOD data to improve the performance of open-set source tracing systems. 

\section{Datasets, Metrics, and Baseline}
\subsection{Dataset and Metrics}
In this work, we adopt the source tracing protocol based on the MLAAD~\cite{muller2024mlaad} dataset~\cite{UsingMLAADforSourceTracing}\footnote{\href{https://deepfake-total.com/sourcetracing}{https://deepfake-total.com/sourcetracing}}. It comprises only spoofed utterances 
and is divided into three partitions: train, development (Dev) and evaluation (Eval). To make the task more challenging, organizers include OOD spoofing classes in the Dev and Eval sets. Table~\ref{tab:MLAAD_split} shows the distribution of ID and OOD classes in each set. We observe an imbalance in the number of samples for certain models on both the ID and OOD sides in each set. Thus for each of the performance metrics we describe in the remainder of this section, we apply class weighting based on the model names of the samples, including those of the OOD data.

We adopt three performance metrics to evaluate open-set performance. For ID classification performance, we report accuracy (\textbf{ID Acc}) on the subset of the Eval data that belong to a seen spoofing model.
The OOD detection performance is evaluated using \textbf{FPR95}~\cite{liu2018open,hendrycksdeep}, which measures the false positive rate of OOD data when the true positive rate of ID data is at 95\%. Additionally, we assess both ID classification and OOD detection performance together using a metric called equal error rate with confusion (\textbf{EERc})~\cite{shon2019mce}\footnote{\href{https://github.com/swshon/multi-speakerID}{https://github.com/swshon/multi-speakerID}}. EERc modifies the standard equal error rate metric to consider ID samples to be correctly classified only if they are correctly classified by both OOD detection \textit{and} ID classification. Lower EERc and FPR95 values indicate better performance.

\begin{table}[t]
\caption{MLAAD source tracing protocol statistics.}
\setlength\tabcolsep{4pt}
\centering
\footnotesize
\vspace{-0.25cm}
\begin{tabular}{@{}lccccc@{}}
\toprule
Set   & ID arch. & ID samples  & OOD arch. & OOD samples \\ \midrule
Train & 24 & 11000   & - & -     \\ 
Dev   & 8 & 4800 & 17 & 7200   \\ 
Eval\tablefootnote{Counts slightly differ from those on the source tracing protocol website. vits--neon samples in the eval set were counted as ``unseen" on the website, despite this system being seen in the train set.}  & 21 & 13591 & 43 & 20309     \\ \bottomrule
\end{tabular}
\label{tab:MLAAD_split}
\vspace{-0.3cm}
\end{table}

\subsection{Baseline}
We adopted a ResNet~\cite{he2016deep} architecture with large margin cosine loss~\cite{wang2018cosface} inspired by its success in~\cite{kleinsource}. However, in contrast with~\cite{kleinsource} that uses ResNet18, we experimented with a wider depth using ResNet34 for further performance improvements. The model consists of
one convolutional layer followed by four residual blocks with depth (3, 4, 6, 3) designed to extract discriminative deep features. A temporal statistics pooling layer (TSPL) is used to aggregate temporal frames to convert the variable-length input to a fixed-length embedding vectors. The TSPL layer computes the channel-wise standard deviation and concatenates it with mean to extract the higher-order statistical features. Finally, a fully-connected linear layer is used to extract 128-dimensional embeddings. The ResNet model uses random 4 second (s) raw audio as input. The front-end features comprise $80$-dimensional log linear filter-bank (LFB) features, extracted from raw audio sampled at $16kHz$, using a $25ms$ window-length and $10ms$ frame-shift. Post-processing includes the extraction of delta ($\Delta$), double delta ($\Delta\Delta$) features, followed by cepstral mean variance normalization (CMVN), resulting in a final feature dimension of $240$. To assess the impact of LMCL, we also implemented a second baseline model that uses the standard softmax cross-entropy loss.

\begin{table}[!]
\centering
\footnotesize
\caption{Baseline ID Acc (\%) $\uparrow$ and OOD detection results.
B:~Baseline w/o LMCL, B+L: Baseline w/ LMCL, E:~Energy.}
\vspace{-0.3cm}
\setlength\tabcolsep{1.4pt}
\footnotesize
\begin{tabular}{ccccccc}
\toprule
\multicolumn{1}{l}{} & \multicolumn{1}{l}{} & \multicolumn{5}{c}{OOD Detection {[}EERc (\%) $\downarrow$ / FPR95 (\%) $\downarrow$ {]}}                                            \\ \cmidrule(){3-7}
           &ID Acc & MSP         & E           & E (T=1/16)  & SME                  & SME (T=4)            \\ \midrule
B & 93.5&18.8/72.9 & 18.5/29.5 & 18.5/29.6 & 18.8/70.3          & \textbf{14.3/20.9} \\
B+L  &93.8& 11.4/14.0 & 41.4/92.3 & 11.4/14.0 & \textbf{10.6/10.9} & 18.8/32.0          \\ \bottomrule
\end{tabular}

\vspace{-0.3cm}
\label{tab:baseline_results}
\end{table}

\subsubsection{Baseline OOD score metrics}
We first assess two popular OOD metrics: maximum softmax probability (MSP)~\cite{hendrycks2022baseline} and energy~\cite{liu2020energy}. Energy is defined as: 
\vspace{-0.3cm}
\begin{align}
E(x;f) &= -T* \log\sum_{i}^{k}{e^{f_i(x)/T}}
  \label{equation:energy}
\end{align}

\vspace{-0.3cm}
where $f_i(x)$ is the real valued logit predicted for the $i$-th class by the model $f$ for sample $x$, and $T$ is the temperature. Both metrics rely on pre-softmax logits for OOD detection. For the baseline model using LMCL, we use the pre-softmax class similarities without applying the margin or scaling so that the scale of the resulting logits is invariant to these parameters.

\subsubsection{Baseline implementation details}
We train the model from scratch for 50 epochs with a batch size of $40$, using a cosine annealing learning rate with an initial value of 1e-3, and a weight decay of 1e-4. To improve generalization, random time and frequency masking augmentations are applied to the LFB features.
The LMCL hyper-parameters are set with minimal tuning based on the Dev set performance, where the scaling factor is set to 16.0, and the margin starts at 0 and linearly increases to reach 0.5 at the 40th epoch.
Experiments on the Dev set confirmed that the energy score could be used without tuning the temperature parameter ($T=1$),
consistent with findings from \cite{liu2020energy}.
However, for the LMCL-based model, tuning the temperature parameter $T$ was needed for optimal performance.
Through minimal tuning on the Dev set, we set $T=1/16$ for the LMCL model. The best model checkpoint was selected based on the lowest EERc on the Dev set, computed using energy as the OOD score metric.

\subsubsection{Initial baseline results}
The ID accuracies of our two baseline models are very similar:
93.5\% without LMCL, and 93.8\%  with LMCL. The OOD detection
performances of baseline models are included in Table~\ref{tab:baseline_results}. For the model without LMCL, energy outperforms MSP, especially FPR95
reducing from 72.9\% to 29.5\%. By using LMCL, we see that we significantly improve performance over the non-LMCL model simply by using MSP where we achieve an EERc of 11.4\% and an FPR95 of 14.0\%. However, for the model with LMCL, using energy with a typical temperature setting of 1.0 yields a significantly degraded OOD detection compared to MSP, with FPR95 increasing to 92.3\%. An explanation for this can be found by observing the average top-class logits predicted by the LMCL model for the ID and OOD data in the Dev set, displayed in Figure~\ref{fig:class_sims}. While the top predicted class logit is noticeably greater for the ID data, the OOD data have greater logits for the second to 10th predicted classes. 
Recall that energy-based OOD detection relies on ID data typically exhibiting greater magnitude energy than OOD data.
Observing equation~\ref{equation:energy}, note that each logit independently contributes to the magnitude of the resulting energy, with larger logits contributing exponentially more. The key difference when utilizing energy with the LMCL model is that the logits are bounded cosine similarities. With this reduced scale of logit, 
the effect of the exponentiation giving more weight to the top logit is reduced, 
allowing the greater 2-10th logits of the OOD data to outweigh the top logit of the ID data. By simply scaling the logits up, we can counteract this effect. Thus, by setting $T = 1/16$, this issue is addressed, leading to very similar results between energy and MSP for the LMCL model.

\section{SME-based Scoring for OOD Detection}

We propose a novel adaptation to the energy score for OOD detection (Eq~\ref{equation:energy}), 
SME, defined as follows:
\vspace{-0.3cm}
\begin{align}
  E_{sm}(x;f) &= -T* \log\sum_{i}^{k}{e^{\sigma_i(f(x)/T)}}
  \label{equation:softmax_energy}
\end{align}

\begin{figure}[!tb]
  \centering
  \includegraphics[clip,width=0.8\linewidth]{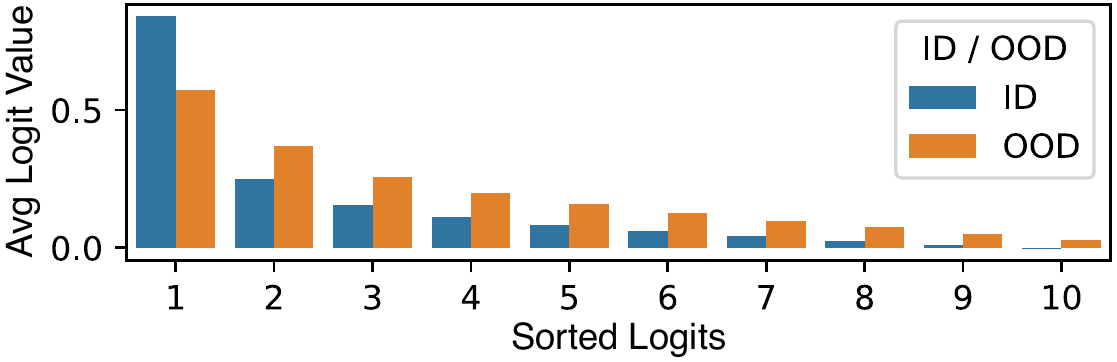}
  \caption{Average top 10 logits for MLAAD Dev set samples, predicted by the LMCL-baseline model.}
  \vspace{-0.55cm}
  \label{fig:class_sims}
\end{figure}

\vspace{-0.3cm}
where $\sigma_i$ denotes the $i$-th output of the softmax function.
Applying softmax on the logits accentuates their skew before computing energy. 
Similar to the effect of scaling the logits as discussed in the previous section, this increases the relative contribution of the top logits to the energy function, helping to increase the energy magnitude for ID samples with more skewed logits than OOD samples (such as those seen in Figure~\ref{fig:class_sims}). 
Additionally, the normalization component of the softmax operation reduces the logits' values if more high valued logits are predicted, leading to a lower energy magnitude. This follows the intuition that an in-domain sample should have a high match score with a small number of classes (ideally 1), since the model was trained for single-label classification.\\
\\
\textbf{Baseline Results.}
For the same baseline models previously evaluated, we compute OOD scores using our newly proposed SME metric and report results in Table~\ref{tab:baseline_results}. Despite the best model checkpoints being chosen using energy, SME yields the best results for the LMCL baseline with an EERc of 10.6\% and an FPR95 of 10.9\%. We find that SME does not provide the same lift with the unbounded logits from the non-LMCL model because the combination of larger scale with softmax results in energy being even further dominated by the top predicted logit, in which case we see SME perform very similarly to MSP. By reducing the scale of the non-LMCL logits before applying SME ($T=4$), we were also able to improve the performance of that model to an EERc of 14.3\% and FPR95 of 20.9\%. Thus, we have found SME to provide value over energy for OOD detection for both LMCL and non LMCL baselines, though we prefer its use with LMCL models in which case scale parameter tuning is not needed.

\begin{table*}[ht!]
\setlength\tabcolsep{10pt}
\centering
\footnotesize
\caption{
Results on MLAAD Eval for models trained with varying auxiliary OOD data. OOD detection is also reported on a subset of OOD data that has speaker overlap with ID train data (Spkr Overlap) and its complement (No Spkr Overlap). Aug: data augmentation.
}
\vspace{-0.35cm}
\begin{tabular}{@{}cccccccc@{}}
\toprule
             &   &       & \multicolumn{5}{c}{\textbf{OOD detection {[}EERc (\%) $\downarrow$ / FPR95 (\%) $\downarrow$ {]}}}                  \\ \cmidrule(l){4-8} 
 & 
   &
   &
  \multicolumn{3}{c}{\textbf{All OOD Data}} &
  \textbf{Spkr Overlap} &
  \textbf{No Spkr Overlap}
  \\ \cmidrule(lr){4-6} \cmidrule(lr){7-7} \cmidrule(lr){8-8} 
\textbf{Aux OOD Data} &
  \textbf{Aug.} &
  \textbf{ID Acc(\%) $\uparrow$} &
  \textbf{MSP} &
  \textbf{E(T=1/16)} &
  \textbf{SME} &
  \textbf{SME} &
  \textbf{SME}
 \\ \midrule
None (Baseline)       & $\times$ & 95.7 & 9.8 / 16.3 & 10.0 / 16.7 & 9.0 / 10.7 &  \textbf{26.2} / 49.0 &  6.7 / 4.5 \\
None (Baseline)     & \checkmark & \textbf{95.8} & 9.7 / 13.6 & 9.7 / 12.8 & 8.8 / 9.9  &  27.0 / 54.0 &  6.0 / 2.7  \\ \midrule

ASVspoof~5    & $\times$ & 95.3 & 9.7 / 14.9 & 10.1 / 15.3 & 8.5 / 9.4  &  28.4 / 48.9 &  6.5 / 2.9  \\
ASVspoof~5-CS & $\times$ & 95.4  & 9.8 / 14.0 & 10.2 / 15.0 & 8.8 / 10.1 &  27.5 / 52.8 &  6.1 / 3.2  \\
\midrule
ASVspoof~5    & \checkmark & 95.6 & 9.9 / 14.1 & 9.9 / 13.6 & 8.8 / 9.6  &  30.1 / 55.2 &  5.7 / 2.2  \\
ASVspoof~5-CS & \checkmark & 95.5 & \textbf{9.2} / \textbf{11.3} & \textbf{9.2} / \textbf{10.9} & \textbf{8.1} / \textbf{8.3}  & 26.5 / \textbf{48.5} &  \textbf{5.5} / \textbf{1.7} \\
\midrule
\multicolumn{2}{c}{\textbf{Average}} & 95.6 & 9.7 / 14.0 & 9.8 / 14.1 & 8.7 / 9.7  & 27.6 / 51.4 & 6.1 / 2.9  \\
\bottomrule
\end{tabular}
\label{tab: OOD results}
\vspace{-0.4cm}
\end{table*}

\section{SME-based Training for OOD Detection}
It has been shown that the OOD detection performance of systems that use MSP or energy can be improved by encouraging separation in the scores for ID and OOD data during training \cite{hendrycks2022baseline, liu2020energy}. \cite{liu2020energy} proposes adding two squared hinge loss terms to the cross-entropy loss to encourage separation by energy score. Inspired by this idea, we adapt the energy loss terms to encourage separation by SME. Our resulting training objective for SME-guided training is
\vspace{-0.1cm}
\begin{align}
  \min_{\theta} \mathbb{E}_{(\mathbf{x_{in}},y)\sim{\mathcal{D}^{train}_{in}}}[\mathcal{L}_{cls}(f(\mathbf{x_{in}}),y)] + \lambda \cdot \mathcal{L}_{Esm} \\
\mathcal{L}_{Esm} = \mathbb{E}_{(\mathbf{x_{in}},y)\sim{\mathcal{D}^{train}_{in}}}(max(0,E_{sm}(\mathbf{x_{in}}) - m_{in}))^2 \\
  + \mathbb{E}_{\mathbf{x_{out}}\sim{\mathcal{D}^{train}_{out}}}(max(0,m_{out} - E_{sm}(\mathbf{x_{out}})))^2
  \label{equation:ood_train_objective}
\end{align}
where $\mathcal{L}_{cls}$ is the original classification loss (e.g. softmax cross-entropy or LMCL), $\mathcal{D}^{train}_{in}$ is the ID training data, $\mathcal{D}^{train}_{out}$ is the auxiliary OOD training data, $\lambda$ is the SME-loss scaling factor, $m_{in}$ is the SME value for ID samples above which a penalty will be incurred, and $m_{out}$ is the SME value for OOD samples below which a penalty will be incurred\footnote{
SME, like energy, is always negative. These thresholds encourage
the SME magnitude to be greater for ID data and lower for OOD data.
}.
\\
\\
We used the following auxiliary OOD datasets:\\
\\
\textbf{ASVspoof~5}: To improve OOD detection, we incorporated the ASVspoof~5~\cite{wang2024asvspoof,wang2025asvspoof} dataset as an auxiliary OOD training data, ensuring that it remains completely disjoint from the MLAAD Eval set. We
selected a balanced subset of samples from each spoofing method and omitted bonafide samples\footnote{We also experimented with including bonafide samples but found worse OOD detection results compared to using only spoofed samples.} to ensure the model focuses on discriminative artifacts of different spoofing methods without being distracted by bonafide speech separation. Along the same lines, we omitted samples that are post-processed with adversarial attacks, since these may be closer to the bonafide speech distribution.

\newpara{ASVspoof 5 copy-synthesis}:
To improve generalization to unseen vocoders and increase diversity in the training data, we generated copy-synthesis (CS) data by re-synthesizing our curated set of spoofed utterances sampled from ASVspoof~5. 
We used five vocoders: Griffin-Lim, Harmonic-plus-Noise Neural Source-Filter (Hn-NSF)~\cite{wang2020neural}, combination of Hn-NSF and HiFiGAN (Hn-NSF-HiFiGAN), WaveGlow~\cite{prenger2019waveglow}, and HiFiGAN~\cite{kong2020hifi}. 
We leveraged pre-trained vocoder models provided in~\cite{wang2023spoofed}. For Griffin-Lim, we used Librosa's~\cite{mcfee2015librosa} implementation\footnote{
We used varying settings such as the  n\_iter (16, 32, 64), hop-length (256, 512, 1024), and window (`hann', `blackman', `hamming').}.

\newpara{Data augmentation strategies}:
We explored codec and reverberation augmentations in improving open-set performance. To introduce real-world encoding variations, Opus and Vorbis codecs were applied with a probability of 0.6, using different bit-rates each time. Reverberation augmentations which introduce acoustic variability and simulate different environments were applied with a probability of 0.4, using randomly selected Room Impulse Response (RIR) and room-size within the range of [0,100]. These augmentation techniques were applied on-the-fly to all training data including auxiliary OOD data used in SME-guided training experiments.

\section{Results}
\subsection{Implementation Details}
For the remaining experiments, we follow the setting of the baseline model with LMCL, except
computing Dev performance using SME.
For fair comparison, we repeat our baseline experiment using SME to compute Dev performance.
For the SME-guided training loss hyper-parameters, we
observed the distribution of SME values that our baseline model produced for ID and OOD data in the Dev set and chose values around the point where the two distributions intersected: $m_{in}=-3.21980$ and $m_{out}=-3.21976$. The scale parameter was then chosen such that the values of the classification loss and SME-loss were similar after the first training epoch: $\lambda=2\text{e}8$. No further hyper-parameter tuning was performed.

\subsection{In-Distribution Classification Results}
The results of our experiments with augmentation and leveraging auxiliary OOD data in training are shown in Table~\ref{tab: OOD results}.
ID accuracy is very slightly compromised when using the auxiliary data sources in training.
We find that the majority of the classification error stems from a small number of models that are challenging to separate. For the baseline, the top two errors are mistaking suno/bark and suno/bark-small for each other ~40\% of the time each. The next most frequent error is mistaking vixTTS for xtts\_v2, which occurs only 3.0\% of the time. 

\subsection{Out-Of-Distribution Detection Results}
Table~\ref{tab: OOD results} also reports the OOD detection performance of our experiments on the MLAAD Eval set under the \textbf{All OOD Data} header. 
We identify a challenging subset of OOD models in the Eval set which trained on the same single-speaker dataset as an ID model. This consists of three LJspeech models, two Thorsten models, and one it/mai\_male model. Performance is reported when limiting OODs to just these six models (\textbf{Spkr~Overlap}) as well as when omitting them (\textbf{No~Spkr~Overlap}).
On the full Eval set, our proposed SME metric outperforms MSP and energy in both EERc and FPR95 for all experiments\footnote{We also experimented with using the original energy loss terms for energy-guided training as well as using energy or MSP to assess Dev performance, however SME consistently gave better results.}, with an average FPR95 of 9.7\%--31\% lower relative to that of scaled energy. Thus for the remainder of this section, we compare experiments using the SME metric.

When applying data augmentation to the baseline, we find moderately improved results on the full Eval set with FPR95 decreasing from 10.7\% to 9.9\%. 
A substantial improvement is seen when omitting the several challenging speaker overlap OOD models, where EERc reduces from 6.7\% to 6.0\% and FPR95 reduces from 4.5\% to 2.7\%. When leveraging OOD data in training, ASVspoof~5 improves performance with FPR95 decreasing from 10.7\% to 9.4\%. ASVspoof~5-CS yields slightly worse results on the full Eval set, though on the subset without speaker overlap we find comparable results: EERc (6.1\% \textit{vs.} 6.5\%) and FPR95 (3.2\% \textit{vs.} 2.9\%) for with-CS and without-CS, respectively.
We find our best results when utilizing codec and reverberation augmentation in addition to leveraging OOD data in training.
Using augmentation and ASVspoof~5-CS, we achieve an EERc of 8.1\% and an FPR95 of 8.3\% on the Eval set, and an EERc of 5.5\% and an FPR95 of 1.7\% when considering only the OOD subset without speaker-overlap.
We find that the variation in artifacts introduced by copy synthesis augmentation is most effective when further augmentations are applied across all train data to mitigate overfitting to the vocoder artifacts.
Finally,
none of our methods made a noticeable improvement on the difficult subset of OOD models that have speaker-overlap with the ID models, where our lowest FPR95 remains 48.5\%\footnote{We attempted to address the challenge with speaker overlap models by using resynthesized ID samples as OOD data in training, however this did not yield an improvement.}.

\section{Discussions}
In this paper, we propose a novel adaptation to the energy score for OOD detection: softmax energy (SME). Replacing the typical temperature-scaled energy score with SME provides a relative improvement in average FPR95 of 31\% across our experiments. Leveraging SME and data augmentation, our best model without the use of auxiliary data achieves an EERc of 8.8\% and an FPR95 of 9.9\%. To compare with the 63\% unweighted EER of the reference system provided by the special session\footnote{\href{https://github.com/piotrkawa/audio-deepfake-source-tracing}{github.com/piotrkawa/audio-deepfake-source-tracing}}, the unweighted EER of our model is 11.3\%, an absolute reduction of 52\%. By utilizing auxiliary data from ASVspoof~5 and copy synthesis, we further improve performance through SME-guided training, yielding 8.1\% EERc and 8.3\% FPR95. Notably, when omitting the 6 models from the Eval set that were trained on the same single speaker datasets as models in our ID set, we achieve an FPR95 of 1.7\%, showing our success on open-set source tracing without this specific challenge. Meanwhile, FPR95 of those 6 models alone is 48.5\%. Thus, an important area for future work is improving the detection of OOD systems that share the same training data as systems in the ID set. Additionally, further experiments are needed to assess the benefit of SME compared to energy with different protocols and models, as well as in different domains such as computer vision.

\bibliographystyle{IEEEtran}
\bibliography{mybib}

\end{document}